\documentclass[12pt]{article}%
\usepackage{amssymb}
\usepackage{amsmath}
\usepackage{amsfonts}
\usepackage{graphicx}%
\providecommand{\U}[1]{\protect\rule{.1in}{.1in}}

\begin{document}
\clearpage

\begin{center}
Hamiltonian Formulation of Ghost Free Mimetic Massive Gravity
\end{center}
\vspace{0.5cm}
\begin{center}
O. Malaeb, C. Saghir \\
Physics Department, American University of Beirut, Lebanon
\end{center}
\vspace{3cm}
\begin{abstract}
We present the hamiltonian formulation of the Ghost Free Mimetic Massive Gravity theory. The linearized theory is studied and the hamitlonian equations of motion are analyzed. Poisson brackets are computed and closure is proved. To prove that this theory is ghost-free, the number of degrees of freedom is analyzed showing that we have only five degrees of freedom. 
\end{abstract}

\clearpage

\section{Introduction}
The idea of Mimetic Gravity was first formulated by Chamseddine and Mukhanov \cite{mimetic} through the parametrization of the physical metric $g_{\mu \nu}$ interms of an auxiliary metric $\tilde{g}_{\mu \nu}$ and a scalar field $\phi$, dubbed mimetic field. The same idea was formulated in \cite{golovnev} without the need of auxiliary metric but through the imposition of an additional constraint on the mimetic field.  In \cite{cosmo}, it was shown that this theory can predict several cosmological solutions: inflation, bouncing universe... through the addition of an arbitrary potential $V(\phi)$ to the action. Dark energy could also be produced through the addition of an extra non dynamical scalar fields associated with new constraints \cite{inhomo}. In \cite{resol} and \cite{black}, the theory of mimetic gravity was extended by the addition of a new function $f(\chi)$, where $\chi= \Box \phi$. Through this function, Chamseddine and Mukhanov were able to prove that both Big Bang and Black Hole singularities are resolved through a specific choice of $f(\chi)$ function. Recently, Chamseddine and Mukhanov \cite{mimass} were able to generate mass of graviton field with a mass term differing form that of Fierz Pauli's one without the presence of an extra ghost field. \paragraph{} 
In this paper we are going to consider the canonical formulation of this theory. The ADM formalism of general relativity was studied by Arnowitt, Deser and Misner in 1959 \cite{dynamics}. The ADM formalism of mimetic theory was constructed in \cite{ola}.  \paragraph{}
The aim of this paper is to analyze the hamiltonian equations of motion and to check that the theory doesn't lead to ghosts. Also, Poisson brackets will be computed.

\section{Canonical Form}
Consider the action of the mimetic massive theory \cite{mimass}:

\begin{align}
I  &  =%
{\displaystyle\int}
d^{4}x\sqrt{g}\left(  -\frac{1}{2}R+\frac{m^{2}}{8}\left(  \frac{1}{2}\bar
{h}^{2}-\bar{h}^{AB}\bar{h}_{AB}\right)  \right. \nonumber\\
& \quad{} \quad \quad \left.  +\lambda\left(  g^{\mu\nu}\partial_{\mu}\phi^{0}\partial_{\nu}%
\phi^{0}-1\right)  \right)  , \label{1}%
\end{align}
where
\begin{align}
\bar{h}^{AB}=g^{\mu\nu}\partial_{\mu}\phi^{A}\partial_{\nu}\phi^{B}-\eta^{AB} \label{2},
\end{align}

To construct the canonical formalism of this theory, the first step is to split the time component from space by rewriting the action in a $3+1$ dimensional form. The mass term and the mimetic one are then given by 
\begin{align}
 S_{\phi} =&{}  \int d^4x  \sqrt{g}  \frac{m^2}{8}  g^{00}\left(-2 -4 \partial_0 \chi^0  -2 \partial_0 \chi^0 \partial_0 \chi^0 -2 \partial_0 \chi^i \partial_0 \chi^k \eta_{ik} \right)  \nonumber \\
& + \sqrt{g}  \frac{m^2}{8} g^{0k} \left( -4 \partial_k \chi^0 -4 \partial_k \chi^0 \partial_0 \chi^0 -4 \eta_{mk} \partial_0 \chi^m -4 \partial_0 \chi^m \partial_k \chi^n \eta_{mn} \right) \nonumber \\
& \sqrt{g}  \frac{m^2}{8} +g^{ik} \left(-2 \eta_{ik} -4 \eta_{mk} \partial_i \chi^m -2 \partial_i \chi^0 \partial_k \chi^0 -2 \eta_{mn} \partial_i \chi^m \partial_k \chi^n \right)  \nonumber \\ 
& + \sqrt{g}  \frac{m^2}{8} g^{00} g^{00}\left(\frac{-1}{2} -2 \partial_0 \chi^0  -3 \partial_0 \chi^0 \partial_0 \chi^0 - \eta_{rs} \partial_0 \chi^r \partial_0 \chi^s \right) \nonumber \\
& +\sqrt{g}  \frac{m^2}{8}  g^{00} g^{0k}\left(-2 \partial_k \chi^0 -6 \partial_k \chi^0 \partial_0 \chi^0 -2 \eta_{rk} \partial_0 \chi^r -4 \eta_{rk} \partial_0 \chi^r \partial_0 \chi^0 -2 \eta_{ir} \partial_0 \chi^i \partial_k \chi^r \right) \nonumber \\
 &+ \sqrt{g}  \frac{m^2}{8} g^{00} g^{ik} \left( \eta_{ik} -\partial_i \chi^0 \partial_k \chi^0 + 2 \partial_0 \chi^0 \eta_{ik} + 2 \partial_i \chi^r \eta_{rk} +4 \partial_0 \chi^0 \partial_i \chi^r \eta_{rk} + \partial_i \chi^r \partial_k \chi^s \eta_{rs} \right. \nonumber \\
& \left. + \partial_0 \chi^0 \partial_0 \chi^0 \eta_{ik} + \eta_{mn} \eta_{ik} \partial_0 \chi^m \partial_0 \chi^n -2 \eta_{mi} \eta_{nk} \partial_0 \chi^m \partial_0 \chi^n -4 \eta_{rk} \partial_0 \chi^r \partial_i \chi^0 \right)  \nonumber \\
& + \sqrt{g}  \frac{m^2}{8} g^{0i} g^{0k} \left( -2 \eta_{ik} -2 \partial_k \chi^0 \partial_i \chi^0 -4 \eta_{ik} \partial_0 \chi^0 -4 \eta_{ri} \partial_k \chi^r -2 \eta_{mr} \eta_{ki} \partial_0 \chi^m \partial_0 \chi^r \right. \nonumber \\
& \left. -2 \eta_{ki} \partial_0 \chi^0 \partial_0 \chi^0 -8 \eta_{ri} \partial_0 \chi^0 \partial_k \chi^r -2 \eta_{rs} \partial_i \chi^s \partial_k \chi^r \right)  \nonumber \\
&  + \sqrt{g}  \frac{m^2}{8} g^{0k}g^{lj} \left( 2 \partial_k \chi^0 \eta_{jl} -4 \eta_{kl} \partial_j \chi^0 +2 \eta_{mk} \eta_{lj} \partial_0 \chi^m -4 \partial_0 \chi^m \eta_{ ml} \eta_{ kj} +2 \partial_k \chi^0 \partial_0 \chi^0 \eta_{lj} \right. \nonumber \\
& \left.  -4 \eta_{kl} \partial_0 \chi^0 \partial_j \chi^0 +4 \partial_k \chi^0 \partial_j \chi^r \eta_{rl} -4 \eta_{rl} \partial_j \chi^0 \partial_k \chi^r -4 \eta_{ks} \partial_j \chi^0 \partial_l \chi^s +2 \eta_{mn} \eta_{lj} \partial_0 \chi^m \partial_k \chi^n \right. \nonumber \\
& \left.  -4 \partial_0 \chi^m \partial_l \chi^r \eta_{mr} \eta_{kj}  -4 \partial_0 \chi^m \partial_k \chi^n \eta_{ml} \eta_{nj} -4 \partial_0 \chi^m \partial_j \chi^s \eta_{ ml} \eta_{ks} +4 \eta_{mk} \eta_{ls} \partial_0 \chi^m \partial_j \chi^s \right)  \nonumber \\ 
& + \sqrt{g}  \frac{m^2}{8} g^{ik} g^{lj} \left( \frac{1}{2} \eta_{ik} \eta_{lj} - \eta_{il} \eta_{kj} +2 \eta_{lj} \eta_{mk} \partial_i \chi^m -4 \partial_i \chi^m \eta_{ml} \eta_{kj} + \partial_l \chi^0 \partial_j \chi^0 \eta_{ik} \right. \nonumber \\
& \left. -2 \eta_{kj} \partial_i \chi^0 \partial_l \chi^0 + \eta_{mn} \eta_{lj} \partial_i \chi^m \partial_k \chi^n  -2 \partial_i \chi^m \partial_l \chi^r \eta_{mr} \eta_{kj} -2 \partial_i \chi^m \partial_k \chi^n \eta_{ml} \eta_{nj} \right. \nonumber \\
& \left.  +2 \eta_{mk} \eta_{rj} \partial_i \chi^m \partial_l \chi^r -2 \eta_{ml} \right)  \nonumber \\
& + \lambda \left( g^{00} \left( 1 +2 \partial_0 \chi^0 +\partial_0 \chi^0 \partial_0 \chi^0 \right)  +2 g^{0k} \partial_k \chi^0 \left( 1 +\partial_0 \chi^0 \right) + g^{ik}\partial_i \chi^0 \partial_k \chi^0 -1 \right)  
\label{4}
\end{align}
Inducing small metric perturbations and considering the small perturbations of the fields around broken symmetry phase, $\phi^A$ and $ g^{\mu \nu}$ can be written as
\begin{align}
\phi^A = x^A + \chi^A
\end{align}
\begin{align}
g^{\mu \nu}= h^{\mu \nu} + \eta^{\mu \nu}
\end{align}
To first order in perturbation, $\bar{h}^{AB}$ becomes
\begin{align}
\bar{h}_{\mu \nu} = -h_{\mu \nu} + \partial_{\mu} \chi_{\nu} + \partial_{\nu} \chi_{\mu} 
\label{coo}
\end{align}
Let us restrict the action to second order in $h_{\mu \nu}$ and $\chi^A$ noting that $\lambda$ is of first order in perturbations. We get the simplified action
\begin{align}
S_{\phi} ={}&  \int d^4x \sqrt{g} \left(\partial_0 \chi^0 \left( \frac{m^2}{8} \left( -2 h^{00} -2 \partial_0 \chi^0 +4 \partial_i \chi^i  + 2 h^{ik} \eta_{ik} \right) + 2 \lambda \right) \right. \nonumber \\
 & \left. + \partial_0 \chi^i \left( \frac{m^2}{8} \left(-4 \partial_i \chi^0 -2 \partial_0 \chi_i -4 h^{0j} \eta_{ij} \right) \right)  \right. \nonumber \\
& \left. + h^{0k} \left( \frac{m^2}{8} \left( -4 \partial_k \chi^0 -2 h^{0i} \eta_{ik} \right) \right) \right.
\nonumber \\
 & \left. + h^{00} \left( \frac{m^2}{8} \left( \frac{-1}{2} h^{00} +2 \partial_i \chi^i + \eta_{ik} h^{ik}\right) + \lambda \right) \right. \nonumber \\ 
& \left. + h^{ik} \frac{m^2}{8} \left( 2 \eta_{ik} \partial^j \chi_j -4 \partial_k \chi_i \right) + h^{ik} h^{lj} \frac{m^2}{8} \left( \frac{1}{2} \eta_{ik} \eta_{lj} - \eta_{il} \eta_{kj} \right) \right. \nonumber\\
& \left.+ \frac{m^2}{8} \left(- 2  \partial_i \chi^0 \partial^i \chi^0 +2 \partial_i \chi^i \partial_k \chi^k 
-2 \partial^k \chi^j \partial_j \chi_k  -2 \partial^i \chi^j \partial_i \chi_j \right) \right) 
\label{5}
\end{align}
The momenta conjugate to $\chi^0$ and $\chi^i$ are respectively  
\begin{align}
P = \frac{\partial L}{\partial \dot{\chi^0}} = \frac{m^2}{8} \sqrt{g} \left( -2 h^{00} +4 \partial_i \chi^i + 2 h^{ik} \eta_{ik} \right) + 2 \sqrt{g} \lambda -4 \sqrt{g} \frac{m^2}{8} \dot{\chi^0}
\label{P} 
\end{align}
\begin{align}
P_i= \frac{\partial L}{\partial \dot{\chi^i}}= \frac{m^2}{8} \sqrt{g} \left( -4 \partial_i \chi^0 -4 h^{0j} \eta_{ij} \right) - \frac{m^2}{2} \sqrt{g}\partial_0\chi_i 
\label{6}
\end{align}
By inverting the above two equations we can write both $\dot{\chi^0}$ and $\dot{\chi^i}$ in terms of their corresponding momenta
\begin{align}
\dot{\chi^0} = \frac{-2}{\sqrt{g} m^2} P + \frac{1}{4} \left( -2h^{00} +4 \partial_i \chi^i+ 2h^{ik} \eta_{ik} \right) + \frac{4 \lambda}{m^2}
\end{align}
\begin{align}
\dot{\chi^i} = \frac{-2 P^i}{m^2 \sqrt{g}}  - \partial_j \chi^0 \eta^{ij} - h^{0i}  \label{7}
\end{align}
In terms of momenta the Hamiltonian becomes
\begin{align}
H_{\phi} {}&= P\dot{\chi^0} + P_i \dot{\chi^i} - L_{\phi}  \nonumber \\
&= - \sqrt{g} \left(h^{ij}\frac{m^2}{8} \left(4 \eta_{ij} \partial_k \chi^k -4 \partial_j \chi_i \right) \right. \nonumber \\
& \left. + \frac{m^2}{8} h^{ij}h^{kl} \left(-\eta_{ik} \eta_{jl} + \frac{1}{2} \eta_{ij} \eta_{kl} \right) \right.\nonumber \\
& \left. +\frac{m^2}{8} \left(-2 \partial_i \chi^j \partial^i \chi^j   - 2 \partial_i \chi^j \partial_j \chi^i +4 \partial_i \chi^i \partial_j \chi^j \right) \right. \nonumber \\
& \left. + \lambda \eta_{ij} h^{ij} +2 \lambda \partial_i \chi^i \right) - \frac{P h^{00}}{2} + \eta_{ij} h^{ij} \frac{P}{2} +P \partial_i \chi^i - \frac{4 \lambda^2 \sqrt{g}}{m^2}  \nonumber \\ 
&+\frac{4 P \lambda}{m^2} - \frac{P^2}{m^2 \sqrt{g}} - \frac{P_i P^i}{m^2 \sqrt{g}} - P_i h^{0i} - P^i \partial_i \chi^0 
\end{align}
The hamiltonian is still a function of the Lagrange multiplier $\lambda$. However, it is independent of its time derivative; therefore, $p_{\lambda} =0 $. This is a primary constraint and will imply a secondary constraint by demanding its time consistency
\begin{align}
0=\dot{p}_{\lambda} = \left \lbrace p_{\lambda}, H \right \rbrace =\frac{\partial H}{ \partial \lambda}.
\end{align}
This allows us to find $\lambda$ and it turns out to be
\begin{align}
\lambda = \frac{P}{ 2 \sqrt{g}} - \frac{m^2}{8} \eta_{ij} h^{ij} - \frac{m^2}{4} \partial_i \chi^i.
\label{lambdaeq}
\end{align}
Substituting back in the Hamiltonian, we end up with
\begin{align}
H_{\phi} ={}&  \int d^4x \left( - \frac{1}{2} P h^{00} - \frac{1}{m^2 \sqrt{g}} P_i P^i - P_i h^{0i} - P^i \partial_i \chi^0  \right. \nonumber \\
& \left. - \frac{m^2}{8}  h^{ij} \left(2 \eta_{ij} \partial_k \chi^k  - 4 \partial_j \chi_i \right) - \frac{m^2}{8}  h^{ij} h^{kl} \left( - \eta_{ik} \eta_{jl} +\frac{1}{2} \eta_{ij} \eta_{kl} \right) \right. \nonumber \\ 
& \left. - \frac{m^2}{8}  \left( -2 \partial_i \chi_j \partial^i \chi^j -2 \partial_i \chi^j \partial_j \chi^i +2 \partial_i \chi^i \partial_k \chi^k \right) \right)
\end{align}

As mentioned above, this hamiltonian is for the mass term and the mimetic one. To form the complete hamiltonian, we still have to add the general relativity part. Considering the Einstein-Hilbert action in terms of the linearized metric $h^{ \mu \nu}$ and keeping it up to second order, we have 
\begin{align}
S_g ={}& - \frac{1}{4} \int d^4x   \left( \partial_{\mu} h^{\mu \nu} \partial_{\nu} h - \partial_{\mu} h^{\mu \sigma} \partial_{\nu} h^{\nu} _{\sigma} + \frac{1}{2} \partial_{\sigma} h^{\mu \nu} \partial^{\sigma} h_{ \mu \nu} + \frac{1}{2} \partial_{\mu} h \partial^{\mu}h \right) 
\label{8}
\end{align}
where $h = \eta^{\mu \nu} h_{\mu \nu}$. Writing this in a $3+1$ dimensional form, it is expressed in terms of the three variables $h^{00}$, $h^{0k}$ and $h^{ik}$. The conjugate momenta corresponding to these variables are \cite{linearized}
\begin{align}
\Pi =  -\frac{1}{4} \partial_m h^{0m} 
\end{align}
\begin{align}
\Pi_l =  \frac{1}{4} \partial_l h^{00} + \frac{1}{4} \partial_l \eta_{rs} h^{rs} - \frac{1}{2} \partial_i \eta_{lr} h^{ri} 
\end{align}
\begin{align}
\Pi_{mn} = \frac{1}{4} \partial_i h^{i0} \eta_{mn} - \frac{1}{4} \eta_{mn} \partial^0 \eta_{rs} h^{rs} + \frac{1}{4} \eta_{mr} \eta_{ns} \partial^0 h^{rs}
\label{PIij}
\end{align} 
By inverting the equation of $\Pi_{mn}$ we can write $\partial_0 h^{qt}$ as follow
\begin{align}
\partial^0 h^{qt} =  4 \eta^{mq} \eta^{nt} \Pi_{mn} -2 \eta^{qt} \eta^{ij} \Pi_{ij} + \frac{1}{2} \partial_i h^{i0} \eta^{qt}
\end{align}
Therefore, the total hamiltonian ($H_{\phi} + H_{g}$) is

\begin{align}
H ={}&  \Pi \dot{h^{00}} + \Pi_{i} \dot{h^{0i}} + \Pi_{ij} \dot{h^{ij}} + P\dot{\chi^0} + P_i \dot{\chi^i} - L  \nonumber \\
& = \int d^4x \left(  +\frac{1}{2} \eta^{ij} \Pi_{ij} \partial_k h^{k0} +2 \Pi_{ij} \Pi_{kl} \eta^{ik} \eta^{jl} - \Pi_{ij} \Pi_{kl} \eta^{ij} \eta^{kl} \right. \nonumber \\
& \left. + \frac{1}{16} \partial_i h^{i0} \partial_j h^{j0}  - \frac{1}{4} \partial_i h^{ij} \partial_j h^{kl} \eta_{kl}  
 + \frac{1}{4} \partial_i h^{ij} \partial_l h^{kl} \eta_{jk} \right. \nonumber \\
& \left. -\frac{1}{4} \partial_i h^{0k} \partial^i h^{0l} \eta_{kl} -\frac{1}{8} \partial_i h^{km} \partial^i h^{ln} \eta_{kl} \eta_{mn} + \frac{1}{8} \partial_i h^{kl} \partial^i h^{mn} \eta_{kl} \eta_{mn} \right. \nonumber \\
& \left. - \frac{1}{m^2 \sqrt{g}} P_i P^i - P_i h^{0i} - P^i \partial_i \chi^0 + h^{00} \left(  \frac{1}{4} \partial_j \partial_i h^{ij} - \frac{1}{4} \partial_i \partial^i h^{kl} \eta_{kl} - \frac{1}{2} P \right) \right. \nonumber \\
& \left. - \frac{m^2}{8}  h^{ij} (2 \eta_{ij} \partial_k \chi^k  - 4  \partial_j \chi_i) - \frac{m^2}{8}  h^{ij} h^{kl} \left( - \eta_{ik} \eta_{jl} +\frac{1}{2} \eta_{ij} \eta_{kl} \right) \right. \nonumber \\ 
& \left. - \frac{m^2}{8}  \left( -2 \partial_i \chi_j \partial^i \chi^j -2 \partial_i \chi^j \partial_j \chi^i +2 \partial_i \chi^i \partial_k \chi^k \right) \right)
\label{bigH}
\end{align}
where $h^{00}$ is appearing as a lagrange multiplier.

\section{Equations of Motion}
Starting from the total hamiltonian constructed above the equations of motion are found by varying with respect to the variables $\chi^0$, $\chi^i$, $P$, $P^i$ and $h^{ij}$.  \newline
The equations of motion of $\chi^0$ and $\chi^i$ are respectively
\begin{align}
\dot{P}={}& - \partial_i P^i
\label{eqP} 
\end{align}

\begin{align}
\dot{P_i} &{}= - \frac{m^2}{4} \partial_i h^{kj} \eta_{kj} + \frac{m^2}{2} \partial_j h^{kj} \eta_{ki} + \frac{m^2}{2} \partial_j \partial^j \chi_i  
\label{eqPm}
\end{align}

The equations of motion of $P$ and $P_i$ are respectively 
\begin{align}
\dot{\chi^0} ={}& -\frac{h^{00}}{2} 
\label{chi0}
\end{align}
\begin{align}
\dot{\chi_i} + \frac{2}{m^2 \sqrt{g}} P_i + \partial_i \chi^0 + h^{0j} \eta_{ji} =0
\label{chiz}
\end{align}
Equation (\ref{chi0}) is exactly the linearized mimetic constraint $\bar{h}^{00} = 0$ 
\begin{align}
\bar{h}^{00} ={}& g^{\mu \nu} \partial_{\mu} \phi^0 \partial_{\nu} \phi^0 -1    \nonumber \\ 
& \left. =  \left(h^{00} +1 \right) \left(1 + \partial_0 \chi^0 \right) \left( 1 + \partial_0 \chi^0 \right) -2 h^{0i} \left(1 + \partial_0 \chi^0 \right) \left(\partial_i \chi^0 \right) \right. \nonumber \\
& \left. + h^{ij} \left(\partial_i \chi^0 \right) \left(\partial_j \chi^0 \right) -1  \right. \nonumber \\
& \left. = h^{00} + 2 \partial_0 \chi^0 = 0 \right.
\end{align}
Substituting for $P_{m}$ in equation (\ref{chiz}) using equation (\ref{6}), and then replacing $h$ by $\bar{h}$, we get
\begin{align}
m^2 \left(\partial^{\rho} \bar{h}_{\rho k} -\frac{1}{2} \partial_k \bar{h} \right)=0
\label{eq1}
\end{align}
The linearized form of the equation $\dot{P}= - \partial_i P^i$ is
\begin{align}
{}& \frac{m^2 \sqrt{g}}{8} \left(-2 \dot{h}^{00} + 4 \partial_0 \partial_i \chi^i +2 \partial_0 h^{ik} \eta_{ik} \right)  \nonumber \\  
& \left. +2 \sqrt{g} \dot{\lambda} -4 \sqrt{g} \frac{m^2}{8} \ddot{\chi}^0 + \frac{m^2 \sqrt{g}}{8} \left( -4 \partial_i \partial_j \chi^0 \eta^{ij} -4 \partial_i h^{0i} \right) \right. \nonumber \\ 
& \left. -\frac{m^2 \sqrt{g}}{2} \partial_i \partial_0 \chi^i = 0 \right.
\end{align}
which proves to be 
\begin{align}
\partial_0 \lambda - \frac{m^2}{4} \left( \partial^{\rho} \bar{h}_{\rho 0} - \frac{1}{2} \partial_0 \bar{h} \right) =0
\label{eq2}
\end{align}
The fourth equation (eq. \ref{chiz}) is simply returning the expression of $P_i$ (eq. \ref{6}). \newline
Using the identity
\begin{equation}
\frac{\partial h^{ij} \left(x \right)}{\partial h^{kl} \left(y \right)} = \frac{1}{2} \left( \delta^i_k \delta^j_l + \delta^i_l \delta^j_k \right) \delta \left(x,y \right),
\end{equation} 
the equation of $h_{ij}$ proves to be 
\begin{align}
G_{ij}\left(- \bar{h}_{\rho \sigma} \right) = - \frac{m^2}{4} \left(\bar{h}_{ij}  - \frac{1}{2} \eta_{ij} \bar{h} \right)
\label{hijeq}
\end{align}
where
\begin{align}
G_{\mu \nu} (h_{\rho \sigma}) ={}& -\frac{1}{2} \left( \partial^2 h_{\mu \nu} - \partial_{\mu} \partial^{\rho} h_{\rho \nu} - \partial_{\nu} \partial^{\rho} h_{\rho \mu} + \partial_{\mu} \partial_{\nu} h \right)  \nonumber \\
& \left. + \frac{1}{2} \eta_{\mu \nu} \left( \partial^2 h - \partial^{\sigma} \partial^{\rho} h_{\rho \sigma} \right) \right.
\end{align}
is invariant under the coordinate transformation (\ref{coo}).\newline
One of the equations of motion obtained is the constraint equation $\bar{h^{00}}$ and the other three equations (\ref{eq1}, \ref{eq2} and \ref{hijeq}) are exactly what was found in \cite{mimass}. 

\section{Poisson Bracket and Degrees of Freedom}

Computing the poisson brackets is important in analyzing the number of physical degrees of freedom. Let us start by considering the primary first class constraints
\begin{align}
T={}&  \Pi + \frac{1}{4} \partial_k h^{0k}   \nonumber \\
T_i ={}&  \Pi_i - \frac{1}{4} \partial_i h^{00} -\frac{1}{4} \partial_i \eta_{rs} h^{rs} + \frac{1}{2} \partial_l \eta_{ir} h^{rl}.
\end{align}
Furthermore, $h^{00}$ appears as a lagrange multiplier. Therefore, we have one more first primary constraint
\begin{align}
N ={}&  -\frac{1}{4} \partial_j \partial_i h^{ij} + \frac{1}{4} \partial_i \partial^i h^{kl} \eta_{kl} + \frac{P}{2}
\end{align}
The time change of these primary first class constraints will lead to a set of secondary frist class constraints. To get the time change, we compute the poisson brackets of the primary constraints with the total hamiltonian.  The time change of T is 
\begin{align}
\left \lbrace T , H \right \rbrace  = + \frac{1}{4} \partial_j \partial_i h^{ij} - \frac{1}{4} \partial_i \partial^i h^{kl} \eta_{kl} - \frac{P}{2} = 0
\end{align}
where $\frac{P}{2}$ is the linearized form of $ \frac{m^2}{8} \sqrt{g} \bar{h} + \lambda \sqrt{g}$.
Thus to first order, the above equation can be written as 
\begin{align}
G_{00} \left(- \bar{h}_{\rho \sigma} \right) = 2 \lambda  + \frac{m^2}{4} \bar{h}
\label{g00}
\end{align}
Equation (\ref{g00}) is equivalent to
\begin{align}
\bigtriangleup \bar{h} + \partial^i \partial^j \bar{h}_{ij} = 4 \lambda + \frac{m^2}{8} \bar{h}
\label{g00prime}
\end{align}
where $\bigtriangleup = - \partial^i \partial_i$. \newline
Similarly, the time change of $T_i$ is
\begin{align}
\left \lbrace T_k , H \right \rbrace ={}&  - \left( + \frac{1}{8} \partial_k \partial_j h^{0j} - \frac{1}{2} \partial_i \partial^i h^{0j} \eta_{kj} - P_k - \frac{1}{2} \partial_k \left( \eta^{ij} \Pi_{ij} \right) \right) \nonumber \\
& \left. + \frac{1}{4} \eta_{ij} \partial_k \left( \dot{h}^{ij} \right) - \frac{1}{2} \eta_{kj} \partial_i \left(\dot{h}^{ij} \right) = 0 \right.
\end{align}
Using the expression of $\Pi_{ij}$ (equation (\ref{PIij})) and that of $P_k$ (equation (\ref{6})), this equation will be given by
\begin{align}
G_{0i}\left(- \bar{h}_{\rho \sigma} \right) = - \frac{m^2}{4} \bar{h}_{0i}
\label{g0i}
\end{align}
which is equivalent to
\begin{align}
\bigtriangleup \bar{h}_{0i} + \partial_0 \partial^k \bar{h}_{ki} + \partial_0 \partial_i \left(\frac{4}{m^2} \lambda  -\frac{1}{2} \bar{h} \right) = m^2 \bar{h}_{0i}.
\label{g0iprime}
\end{align}
The above equations (\ref{g00}, \ref{g00prime}, \ref{g0i} and \ref{g0iprime}) are exactly those found in \cite{mimass}. \newline
The time change of N is
\begin{align}
\left \lbrace N , H \right \rbrace ={}&  0 
\end{align}

Counting the degrees of freedom, we have ten independent fields $h_{\mu \nu}$ and four independent field $X^A$. This will give us a total of fourteen degrees of freedom. There are four primary first class constraints which lead to a set of four secondary class constraints. This will leave us with six degrees of freedom. There is one additional primary first class constraint; therefore, we end up having five degrees of freedom representing the massive graviton. 
\section{Looking at the Mimetic term}
The hamiltonian (eq. \ref{bigH}) appears to be independent of lambda. To investigate the energy density of the mimetic term,this hamiltonian must be expanded up to second order in scalar perturbations \cite{long}. For small perturbations, different fields are expanded as follows
\begin{align}
& \left. \chi^0=\chi^0 \right. \nonumber \\
& \left. \chi^i = \tilde{\chi}^i - \partial^i \pi \right. \nonumber \\
& \left. h^{00}= -2 \phi \right. \nonumber \\
& \left. h^{0i}=0 \right. \nonumber \\
& \left. h^{ij}= 2 \psi \eta^{ij} \right.
\label{perturbation}
\end{align}
Due to the mimetic constraint $\bar{h}^{00}=0$, we get $h^{00}=- 2 \dot{\chi}^0$.
Substituting these pertubations in (eq. \ref{bigH}), the  scalar part of the hamiltonian becomes 
\begin{align}
H_{scalar} ={}&  \int d^4x  \left( \psi \bigtriangleup \psi -\frac{3m^2}{4}\psi^2 -2\dot{\chi^0} \bigtriangleup \psi - \frac{m^2}{2} \psi \bigtriangleup \pi  + \frac{m^2}{4} \bigtriangleup \pi \bigtriangleup \pi \right. \nonumber \\ 
& \left. + \dot{\chi^0} P - P^i \partial_i \chi^0 - 3\dot{\psi^2} - \frac{1}{m^2}P_i P^i \right)
\label{hasca}
\end{align}
Varying the above hamiltonian with respect to $\chi^0$ where the momenta (eq.\ref{P} and eq.\ref{6}), up to first order in scalar perturbations, are respectively
\begin{align}
& \left. P= 2 \bigtriangleup \psi \right. \nonumber \\
& \left. P_i = -\frac{m^2}{2} \partial_i \chi^0 + \frac{m^2}{2} \partial_i \dot{\pi} \right.
\end{align}
We get
\begin{align}
\dot{\psi}=\frac{-m^2}{4}\left( \chi^0 - \dot{\pi} \right)
\label{scalar}
\end{align}
which is exactly the scalar perturbation of (eq \ref{eqP})
\begin{align}
\dot{P}=-\partial_i P^i
\end{align}
\par

In terms of $\psi$ and $\pi$, the hamiltonian eq \ref{hasca} becomes
\begin{align}
 H_{scalar} ={}&  \int d^4x \left(  \psi \left( \bigtriangleup -\frac{3 m^2}{4} \right) \psi + \frac{4}{m^2} \dot{\psi} \left( \bigtriangleup - \frac{3 m^2}{4}\right) \dot{\psi} \right. \nonumber \\
& \left. - \frac{m^2}{4} \left(2 \psi \bigtriangleup \pi - \bigtriangleup \pi\bigtriangleup \pi \right) - 2 \dot{\pi} \bigtriangleup \dot{\psi} \right)
\label{hascal}
\end{align}
To get rid of the mixed terms, we need to diagonalize the above hamiltonian. Starting from the equation of motion of $h_{00}$ (eq. \ref{g00prime}) or from the equation of lambda (eq. \ref{lambdaeq}) we get the equation of motion of $\psi$
\begin{align}
\bigtriangleup \psi -\frac{3 m^2}{4} \psi = \lambda + \frac{m^2}{4}\bigtriangleup \pi
\end{align}
which gives
\begin{align}
& \left. \psi= \left(\bigtriangleup - \frac{3 m^2}{4} \right)^{-1} \left( \lambda + \frac{m^2}{4} \bigtriangleup \pi \right)\right.
\end{align}

Substituting $\psi$ in (eq. \ref{hascal}), the hamitlonian for the pure mimetic term becomes  
\begin{align}
H_{\lambda} = {}&  \int d^4x \left(\frac{-16 \dot{\lambda}^2 }{m^2 \left(3 m^2 - 4 \bigtriangleup \right)}  - \frac{4 \lambda^2}{ \left(3 m^2 -4\bigtriangleup \right)}  \right)
\label{Hlambda}
\end{align}
knowing that the corresponding momentum $p_\lambda$ of $\lambda$ is
\begin{align}
p_{\lambda}= \frac{-32\dot{\lambda}}{m^2 \left(3m^2- 4 \bigtriangleup \right)}
\end{align}
eq. \ref{Hlambda} becomes
\begin{align}
H_{\lambda}= -\frac{p_{\lambda}^2 m^2 \left(3m^2 -4 \bigtriangleup \right)}{64} - \frac{4 \lambda^2}{3m^2 - 4 \bigtriangleup}
\end{align}
 For plane wave modes of wave number $ \vec{k}$, $\bigtriangleup$ is $-k^2$
For modes with $ k \gg m$, the above energy density reduces to
\begin{align}
-\frac{-4}{m^2 k^2} \left( \dot{\lambda ^2} + \frac{m^2}{4} \lambda^2 \right)
\end{align}
The propagtor of the $\lambda$ term shows that $\dot{\lambda} \propto m \lambda$. This fact avoid the singularity $m^2 \rightarrow 0$.
For $m^2 \rightarrow 0$ the energy density of the $\lambda$ term becomes
\begin{align}
\epsilon_{mim} \simeq \lambda - \frac{\lambda^2}{k^2}
\end{align}
For $\lambda \ll k^2$, the energy density is positive and the linear term dominats. One should not worry about the other case where $\lambda > k^2$ since perturbation theory will be no more valid in this limit.

\section{Conclusion}
In this paper, we have constructed the Hamiltonian formulation of Ghost Free Mimetic Massive gravity. In their approach, small metric perturbations were induced small perturbations of the fields around broken symmetry phase were considered. Therefore, the Einstein-Hilbert action was expressed in terms of the linearized metric $h^{\mu \nu}$. \newline
We wrote the total action in a $3+1$ dimensional form up to second order in perturbations, and found the momenta of each field. This enabled us to write the total Hamiltonian.  The equations of motion were all found and it was proved that they are the same equations obtained in \cite{mimass}. At the end, the poisson brackets of the contraints were computed and degrees of freedom were counted proving that we have five degrees of freedom corresponding to a massive graviton without ghosts.

\clearpage

\textbf{{\large {Acknowledgments}}}

We would like to thank Professor Ali Chamseddine for suggesting the problem and for his helpful discussions on the subject. We would like to thank also the American University of Beirut (Faculty of Science) for the support.

\bigskip

\bigskip

\clearpage

\begin{thebibliography}{9} %

\bibitem {mimetic} \textsc{A. H. Chamseddine and V. Mukhanov}:\ \textit{Mimetic Dark Matter}, JHEP, \textbf{1311}, (2013), 135.

\bibitem {golovnev} \textsc{A. Golovnev}:\ \textit{On the recently proposed Mimetic Dark Matter}, Physics Letters B, \textbf{728}, (2014), 39.

\bibitem {cosmo} \textsc{A. H. Chamseddine, V. Mukhanov, A. Vikman}:\ \textit{Cosmology with Mimetic Matter}, JCAP 1406 (2014) 017, 19 pp.
\bibitem {inhomo} \textsc{A. H. Chamseddine and V. Mukhanov}:\ \textit{Inhomogeneous Dark Energy}, JCAP 1602 (2016) no.02, 040, 11 pp.
\bibitem {resol} \textsc{A. H. Chamseddine and V. Mukhanov}:\ \textit{Resolving Cosmological Singularities}, JCAP 1703 (2017) no.03, 009, 15 pp.
\bibitem {black} \textsc{A. H. Chamseddine and V. Mukhanov}:\ \textit{Nonsingular Black Hole}, Eur.Phys.J. C77 (2017) no.3, 183 , 21 pp.
\bibitem {mimass} \textsc{A. H. Chamseddine and V. Mukhanov}:\ \textit{Ghost Free Mimetic Massive Gravity}, JHEP 1806 (2018) 060, 3pp.
\bibitem {dynamics} \textsc{R. Arnowitt, S. Deser and C. W. Misner}:\ \textit{Republication of: The dynamics of general relativity}, General Relativity and Gravitation, \textbf{40}, (2008), 1997.
\bibitem {ola} \textsc{O. Malaeb}:\ \textit{Hamiltonian Formulation of Mimetic Gravity}, Phys. Rev. D 91, 103526 (2015), 15 PP.
\bibitem {long} \textsc{A. H. Chamseddine and V. Mukhanov}:\ \textit{Mimetic Massive Gravity: Beyond Linear Approximation}, JHEP 1806 (2018) 062و 25 pp.
\bibitem {linearized} \textsc{R. N. Ghalati}:\ \textit{Constraint Analysis of Linearized Gravity and a Generalization of the HTZ Approach}, arXiv: hep-th/0703268v1

\end{thebibliography}
\end{document}